\input{aipcheck}

\documentclass[
    ,final            
  ]
  {aipproc}

\layoutstyle{6x9}

\newcommand{\glog}{\Lambda}
\newcommand{\gexp}{{\cal E}}

\newcommand{\ep}{\epsilon}

\begin{document}

\title{Entropies for complex systems: generalized-generalized entropies}


\author{Stefan Thurner$^{1,2}$}{
}

\author{Rudolf Hanel$^1$}{
  address={$^1$Complex Systems Research Group, HNO, 
   Medical University of Vienna, W\"ahringer G\"urtel 18-20, A-1090, Austria \\
   $^2$Santa Fe Institute, 1399 Hyde Park, Santa Fe, NM 87501, USA }
}

\begin{abstract}
 Many complex systems are characterized by non-Boltzmann distribution functions of their statistical 
variables. If one wants to -- justified or not -- hold on to the maximum entropy principle for complex statistical 
systems (non-Boltzmann) we demonstrate how the corresponding entropy has to look like, given the form of the 
corresponding distribution functions. By two natural assumptions that (i) the maximum entropy principle 
should hold and that (ii) entropy should  describe the correct thermodynamics of a system (which 
produces non-Boltzmann distributions) the existence of  a class of fully consistent  entropies can be deduced. 
Classical Boltzmann-Gibbs entropy is recovered as a special case for the observed
distribution being the exponential, Tsallis entropy is the special case for q-exponential observations.
\end{abstract}

\maketitle


\section{Introduction}

It has been realized that many statistical systems in nature can not 
be described by a naive or straight forward application of Boltzmann-Gibbs 
statistical mechanics. 
In contrast to ergodic, separable, locally and weakly interacting systems, 
these systems are {\it complex} systems whose characteristic distributions 
often are of power-law type, or more complicated. 
Due to the existence of strong correlations between its elements complex systems 
often violate ergodicity and are prepared in states at the {\it edge of chaos}, 
i.e. they exhibit weak sensitivity to initial conditions. 
Further, complex systems are mostly not separable in the sense, that probabilities 
for finding a system in a given state factorize into single particle probabilities 
and as a consequence, renders these systems not treatable with Boltzmann single particle 
entropies \cite{ludwig}. 
In this context an adequate starting ground is provided by Gibbs entropies
\begin{equation}
 S[B]=
  - \int d\Gamma \,\, B \left(H(\Gamma)\right) \,\, 
  \log\left(B\left(H(\Gamma)\right)\right) \quad ,
\label{gibbs}
\end{equation} 
where $\Gamma$ are the phase space variables, and $B$ is the 
distribution function (Boltzmann factor) and $[B]$ indicates functional dependency on $B$. 
Gibbs entropies can in principle incorporate arbitrary correlations via their explicit
dependence on the Hamiltonian function $H(\Gamma)$ (potential term) describing the system. 
However, the Gibbs entropy in combination with the usual Jaynes maximum ignorance principle 
(maximum entropy principle or variational principle) \cite{jaynes} usually fixes the distribution 
function to be of exponential type $B(H)\sim \exp(-\beta H)$, as 
demonstrated
in detail below. 
For extending the concept of statistical mechanics to complex systems, 
which are characterized by fundamentally different distribution functions, it becomes  
necessary to consider generalizations of the exponential 
distribution function. 
It is interesting to note that the exponential form of the 
distribution function is not a priori dictated by classical statistical mechanics, 
but in contrast  much of classical statistical mechanics is built upon this special form 
of the distribution function, as argued e.g. in \cite{kaniadakislog}. It is possible, e.g., to 
construct non-exponential distributions for particle systems. The form of these distributions
depend on the form of the inter-particle potentials \cite{hanel1, hanel2}. 

The aim here is to construct and deduce a correct entropy starting from a given experiment
on an arbitrary statistical system \cite{hanelthurner2007}. Given a measured distribution function, 
(e.g. experimental data), what is the associated entropy which is compatible with the 
maximum ignorance principle? 
This philosophy is very different from what has been done so far, 
i.e. take a (possibly modified) entropy and understand the resulting distribution functions.

\section{Entropies for complex systems ?}

In the following we ask whether one can construct a self-consistent theoretical framework
where data, i.e. the measured distribution, serves as a starting point 
to construct an entropy which is consistent with both, the 
correct thermodynamic relations and the Jaynes maximum entropy principle \cite{jaynes}.
According to this modification of logics it is 
sensible, in a first step, to modify or deform the $\log$ in Eq. (\ref{gibbs}) 
to a generalized logarithm $\Lambda$.
The concept of deforming logarithms and thus modifying the form of entropy to accommodate 
a large body of experimental data from complex systems is not new
\cite{kaniadakislog,tsallis88,tsallis05,abelog,naudts,wada,borges,filho}.
The generalized Gibbs entropy then reads
\begin{equation}
    S_G[B]=-\int d\Gamma \,\,
				B\left(H\right)\glog\left(B\left(H\right)\right) \quad ,				
\label{gge}
\end{equation}
which, by a simple computation, performing the $\Gamma$ integration over energy spheres, can be rewritten to 
\begin{equation}
   S_G[B]=
      - \int d\ep  \,\, \omega(\ep) \,\, 
				B\left(\ep\right)\glog\left(B\left(\ep\right)\right) \quad ,
\label{gge_alt}
\end{equation}
where $\omega(\ep) \equiv \int d\Gamma \delta(E-H)$ 
is the microcanonical multiplicity factor for the total system energy $E$. 
The associated maximum ignorance principle states that distribution functions 
of statistical systems are obtained by maximizing  a functional $G$
\begin{equation}
  G =  S_G[B] - \beta \int d\ep \,\, \omega\left(\ep\right) B 
\left(\ep\right)(\ep-U)- \gamma \left( \int d\ep \,\, \omega\left(\ep\right)  B(\ep) - 1\right)   \quad ,
\label{stdvarprinciple}
\end{equation} 
provided the knowledge (measurement) of the expected energy value $U$ alone.
Here $\beta$ is the usual inverse temperature
and $\gamma$ is the Lagrange parameter for normalizability.
To keep close formal contact with usual statistical physics, we now 
represent the measured distribution function
by replacing the usual exponential function by some positive 
function $\gexp$, i.e.
\begin{equation}
  \exp(-\beta(E-U)-\tilde{\gamma}) \to \gexp(-\beta(E-U)-\tilde{\gamma}) \quad ,
\label{gen_boltz_fac}
\end{equation}
where $\tilde{\gamma}$ is the normalization constant. 
 $\gexp$ is the {\em deformed exponential} function. 
It is the inverse function of $\glog$.
The existence of the inverse function requires $\gexp$, and thus $\glog$, to be monotonous.
The minimum requirements for a distribution function $B$ are that 
(i) $B$ is positive and monotonic; 
(ii) $B$ can be normalized, i.e. $\int d\ep\, \omega(\ep)B(\ep)=1$; 
(iii) $B$ is sufficiently stable in time, 
such that it can be seen  as a reasonable measurable probability. 
The generalized Gibbs entropy Eq. (\ref{gge_alt}) then reads 
\begin{equation}
 S_G[B]= \int d\ep  \,\, \omega(\ep) \,\, 
        \gexp\left(-\beta (\ep-U)-\tilde{\gamma}\right) 
      \left(\beta (\ep-U)+\tilde{\gamma}\right) \quad ,
\label{ge}
\end{equation}
and the usual definition of the expectation value 
\begin{equation}
\left<f\right> \equiv 
\int d\ep \, f(\ep) \, \omega(\ep) \, \gexp\left(-\beta (\ep-U)-\tilde{\gamma}\right)
\end{equation}
holds. Obviously the normalization constant $\tilde{\gamma}$ has to be chosen such that 
\begin{equation}
  \int d\ep \,\, \omega(\ep) \,\,
  \gexp\left(-\beta (\ep-U)-\tilde{\gamma}\right) = 1\quad.
  \label{normalization_A}
\end{equation}

\subsection{A problem}

However, this approach raises a problem.
Variation, i.e.,  $\delta G=0$ with respect to $B$ obviously implies
\begin{equation}
\frac{d}{dB}B\glog\left(B\right)=-\gamma-\beta\left(E-U\right) \quad . 
\label{problem_var}
\end{equation}
With the desired form of the distribution function,  
 $B(E)=\gexp(-\beta(E-U)-\tilde{\gamma})$
the only solution 
for the generalized Gibbs entropy Eq. (\ref{gge_alt}) is the logarithm (!), i.e. $\glog(B)\propto\log(B)$. 
This is because Eq. (\ref{problem_var}) rewrites into
$\glog(B)+B\glog(B)'=-\gamma-\beta\left(E-U\right)$. Inserting $B(E)=\gexp(-\beta(E-U)-\tilde{\gamma})$ 
into $\glog(B)$ further implies $B\glog(B)'=\tilde\gamma-\gamma=\mbox{const.}$, which in turn implies,  
$\glog(B)'=\mbox{const.}/B$.
Thus $\gexp$ is forced to represent the usual exponential statistics.
This is completely unsatisfactory for our philosophy!

\subsection{Solving the problem}

The above problem arises because of the extra term in Eq. (\ref{problem_var}), $B\glog'(B)$, 
which is non-trivial for the $\glog$ being anything other than the $\log$.  
To cancel this term we suggest to further generalize the generalized logarithm $\glog(B)$ to a {\em functional} 
\begin{equation}
\glog(B) \rightarrow \bar \glog[B]\equiv   \glog(B) - \eta[B] \quad ,
\end{equation} 
where $[ B ]$ again indicates functional dependence on $B$. 
By substituting  $\glog$ by $\bar \glog$ in Eq. (\ref{gge}), we obtain the 
{\em generalized-generalized} entropy 
\begin{equation}
S_{GG}[B] \equiv   S_G[B] + \eta[B] \quad ,
\label{entr}
\end{equation}
where we have used that $\eta$ is a constant with respect to $\epsilon$-integration and 
the normalization condition, Eq. (\ref{normalization_A}). 
Now, the idea is that after variation with respect to $B$, the additional term 
$\frac{\delta}{\delta B} \eta[B]$ exactly cancels the problematic term, $B\glog'(B)$, ore more precisely,
 $-\omega(E) B(E) \frac{d}{dB} \glog(B(E))$. 
The corresponding condition, $\frac{\delta}{\delta B} \eta[B]=\omega(E) B(E) \frac{d}{dB} \glog(B(E))$, 
now dictates the form of $\eta$ 
\begin{equation}
\eta[B] = \int d\ep \,\, \omega(\ep) \int_0^{B(\ep)} dx \,\, \glog'(x) x   + c \quad ,
\label{ansatz}
\end{equation}
up to an integration constant $c$. By partial integration of Eq. (\ref{ansatz}) 
and substituting the result into Eq. (\ref{entr}) 
we get for the  generalized-generalized  entropy 
\begin{equation}
  S_{GG}[B] =  -\int d\ep \,\, \omega(\ep) \int_0^{B(\ep)} dx \,\, \glog(x)  +  \bar c\quad ,
\label{newent}
\end{equation}
with $\bar c$ an integration constant which is only different from $c$, iff
$\lim_{x\to 0} x\glog(x) \neq 0$. This constant $\bar c$ can be fixed to impose $S_{GG}=0$ for the completely
ordered state.  
Note that,  based on a purely thermodynamic argument \cite{weiner}, 
a very similar form of an entropy has been derived in \cite{almeida}. 
In a very different context the same function was observed in \cite{naudts2004}. 

It can now easily be checked that this entropy, Eq. (\ref{newent}), in combination with 
the standard maximum ignorance principle under the {\it usual} constraints,   
yields the measured distributions $B$. Define
\begin{equation}
 G = S_{GG}[B] - \beta  \int d\ep \,\, \omega(\ep) \,\, B(\ep) \,\, (\ep-U)
          - \gamma\, \left( \int d\ep \,\, \omega(\ep) \,\, B(\ep)-1 \right)  \quad , 
\label{funct2}
\end{equation}
and vary with respect to $B$, $\frac{\delta}{\delta B}G=0$. Dividing by $\omega(E)$ yields 
\begin{equation}
B(E) \glog'(B(E))- \gamma  - \beta(E -U) - \frac{d}{d B} B(E) \glog(B(E)) =0 
\quad , 
\label{vari}
\end{equation}
or equivalently, $\glog(B(E))=-\gamma-\beta(E-U)$. Using that 
$\gexp$ is the inverse of $\glog$, the correct generalized
distribution function,   $B(E)=\gexp(-\beta(E-U)-  \gamma)$,   is recovered. 
The normalization constant $\gamma=\tilde\gamma$  is equivalent with the 
Lagrange parameter of the maximum entropy functional.  

\section{Thermodynamics of generalized-generalized entropies}

To show that the proposed entropy of Eq. (\ref{newent}) is fully consistent with 
the necessary thermodynamic relations we first   
differentiate Eq. (\ref{newent}) with respect to $U$ to get
$-\int d\ep\,\omega(\ep)\glog(B(\ep))\partial B(\ep)/\partial U$.
Then, substituting $-\gamma-\beta(E-U)$ for $\glog(B(E))$ and 
partially integrating the result, taking into account that both,  
$\gamma$ and $\beta$ depend on $U$, leads us to the (generic)  result
\begin{equation}
 \frac{\partial}{\partial U} S[B]= \beta   \quad . 
\end{equation}
The first and second laws for the generalized-generalized entropy are 
reviewed from \cite{abe07}, where the entropy functional Eq. (\ref{newent}) 
is written in its discrete form
\begin{equation}
S_{GG}[B] = - \sum_i  \int_0 ^{B(\epsilon_i)}  dx\,\, \Lambda(x) + \bar c \quad .
\label{newdiscreteent}
\end{equation}

\subsection{The first law}

Since we use the usual concept of  measurement, internal energy is given by 
$U=\sum_i \epsilon_i B(\epsilon_i)$. Let us vary $U$ to get
\begin{eqnarray}
   \delta U = & \underbrace{\sum_i \epsilon_i  \delta B(\epsilon_i) } & 
  + \underbrace{ \sum_i B|_{B=\tilde B} \delta \epsilon_i }  \quad ,
                      \nonumber \\ 
      & \bar \delta Q  &   \qquad   \quad - \bar \delta W
\end{eqnarray}
because first
\begin{equation}
 \delta S_{GG} = -\delta  \sum_i \int_0^{B(\epsilon_i) }  dx\,\, \Lambda (x) |_{B=\tilde B} 
= \beta \sum_i \epsilon_i \delta B \equiv  \beta  \bar \delta Q \quad , 
\end{equation}
where $Q$ denotes heat, and second,  work is identified as usual as 
\begin{equation}
\delta W \equiv - \sum_i B|_{B=\tilde B} \delta \epsilon_i  \quad . 
\end{equation}

\subsection{The second law}

Let us write the {\em relative} entropy associated with the entropy of 
Eq. (\ref{newdiscreteent}), which is of Bregman type, 
 \begin{equation}
    K_{\Lambda } \left[  B || \bar B \right] = \sum_i \int_{\bar B(\epsilon_i ) }^{B(\epsilon_i)} dx\,\, ( \Lambda (x)     
     -\Lambda(\bar B)  ) \quad . 
 \end{equation}
Its variation with respect to $B$ yields
 \begin{equation}
    \delta K_{\Lambda } \left[  B || \bar B \right] = \sum_i [\Lambda(x)-\Lambda(\hat B)  ] \delta B(\epsilon_i) \quad .
 \end{equation}
Now, take a maximum-entropy distribution, $\hat B =\Lambda ^{-1}(\gamma-\beta (\epsilon_i -\hat U) )$ 
as the reference state
 \begin{equation}
    \delta K_{\Lambda } \left[  B || \bar B \right] = \sum_i     \Lambda(x) \delta B(\epsilon_i)   + \beta \sum_i \epsilon_i 
\delta B(\epsilon_i)  = -\delta S_{GG} + \beta \delta Q \quad . 
\end{equation}
Supposing the existence of   a third state, $B^*$, with
 \begin{equation}
   K_{\Lambda } \left[  B^* || \bar B \right]  \leq K_{\Lambda } \left[  B || \bar B \right] \quad , 
 \end{equation}
one can write the variation as a superposition 
 \begin{equation}
   \delta K_{\Lambda } \left[  B || \bar B \right]  = K_{\Lambda }\left[ \lambda B^* + (1-\lambda ) B || \bar B \right]   -  K_{\Lambda }\left[  B || \bar B \right]  
 \quad , \quad  \forall \lambda \in (0,1) \quad .
\end{equation}
Since relative entropy is convex, 
\begin{equation}
 K_{\Lambda }\left[ \lambda B^* + (1-\lambda ) B || \bar B \right]   -  K_{\Lambda }\left[  B || \bar B \right]   \leq
\lambda K_{\Lambda }\left[  B ^* || \bar B \right]   + (1-\lambda ) K_{\Lambda }\left[  B || \bar B \right] 
\end{equation}
and 
\begin{equation}
  \delta K_{\Lambda }\left[  B || \bar B \right] = \lambda \left(   K_{\Lambda }\left[  B^* || \bar B \right] -K_{\Lambda }\left[  B || \bar B \right]  \right) \leq 0
  \quad , 
\end{equation}
this means that the Clausius inequalities hold for 
generalized-generalized entropies, i.e. 
\begin{equation}
  \delta S_{GG} \geq \beta \delta Q \quad .
\end{equation}

\section{Classical Examples}

Let us give two examples to illustrate our philosophy, now switching back to continuous notation. 
{\it Example 1: Classical  Boltzmann distributions.} 
If the experimentally measured distribution is of exponential type,
$B(E)\sim \exp(-\beta E)$, then $ \glog (B) \sim \ln(B)$,
and using Eq. (\ref{newent}) yields the usual Boltzmann entropy, 
\begin{equation}
S_{GG}[B]= -\int d\ep \,\, \omega(\ep) \,\, B(\ep) \,\, \ln(B(\ep)) +1 +\bar c \quad .
\end{equation}
\\
{\it Example 2: Asymptotic power-law distributions.}
Let us suppose an experimental measured distribution function is a $q$-exponential, i.e.  
$B(E)=\left[1-(1-q) E \right]^{\frac{1}{1-q}}$. Thus, the generalized logarithm is the 
so-called $q$-$\log$, 
$\glog(B) =  \ln_{q}(B) \equiv  \frac{B^{1-q}-1}{1-q} $.
Inserting as before gives the Tsallis entropy \cite{tsallis88,tsallis05} times a factor,
\begin{equation}
   S_{GG}[B] = - \frac{1}{2-q}  \int d\ep \,\, \omega(\ep) \,\, B(\ep) \,\, \ln_q(B(\ep)) +\frac{1}{2-q}  +\bar c
   \quad ,
\end{equation}
where we require $q<2$. The factor can in principle be absorbed into a transformation 
of $\beta$ and $\gamma$.
At this point it is also obvious that in the case of power law distributions 
the question of normalizability can arise.
Note however, that since $\rho=\omega B$ has to be normalizable,
an implicit regularization is provided by
the maximal energy $E_{\rm max}$ the observable system (represented by $\omega$)
can assume.

\section{Generalized distributions and dual logarithms}

It has been recently noted \cite{wada,naudts,borges,tsallis05} that the generalization of distribution functions
immediately involves the existence of {\it dual logarithms} $\glog^*(x) \equiv -\glog(1/x)$, as has first been defined in 
\cite{naudts_physA2002}.
For the generalized Gibbs entropy Eq. (\ref{gge}), for instance, one only has to consider the energy constraint 
$\left<\ep\right>=U$ together with Eq. (\ref{ge}) to find $S_G[B]=\tilde{\gamma}$. 
For $\beta=0$, this implies that $B(E)=Z^{-1}=\mbox{const}$, because $Z=\int d\ep \,\, \omega(\ep)$. Therefore,
$S_G[B]=-\int d\ep \,\, \omega(\ep)Z^{-1}\glog(Z^{-1})   =   -\glog(Z^{-1})=\glog^*(Z)$. 

However, the dual logarithm property can be made genuine for arbitrary values of
$\beta$ and generalized logarithms within the setting of our proposed generalized-generalized 
entropy Eq. (\ref{newent}), by
observing that the partition function $Z$ has to be defined in a deformed way, too. 
Using the definition of the deformed product 
$x\otimes y \equiv \gexp(\glog(x)+\glog(y))$, as given in \cite{kaniadakislog},
the renormalization condition for the distribution function $B$ can then be recast into the form
\begin{equation}
 B(E)=
 \left(\frac{1}{Z}\right)\otimes
 \gexp\left(-\beta \left(E-U\right)\right) \quad , 
 \label{boltz2}
\end{equation}
which is to say, $\gamma=-\glog(1/Z)$ or $Z=1/\gexp(-\gamma)$.
This becomes the defining equation for the generalized partition function $Z$. 
Inserting this into Eq. (\ref{newent}) we finally get 
\begin{equation}
 S_{GG}= \eta + \glog^*(Z) \quad ,
 \label{gendual}
\end{equation}
generalizing the famous formula $S=\log(Z)$ and giving us a deeper insight on
how the functional $\eta$ mediates between entropy and partition function.
For self-dual logs, $\glog^*\equiv\glog$, which excludes the 
$q$-logarithm ($\ln_{q}^{*}(p)=\ln_{2-q}(p)$) and the Abe-log \cite{abelog},
Eq. (\ref{gendual}) reduces to, $S_{GG}= \eta+\glog(Z)$.

\section{Discussion}

We start by relaxing the restriction that distribution functions 
have to be exponentials and allow arbitrary types of distributions, $B(E)$. 
By doing so we introduce corresponding generalized logarithms, $\glog$, the 
inverses of $B$, and suggest 
to construct the entropy of systems leading to non-exponential distributions, 
as $S=-\int d\ep \,\, \omega(\ep) \int_0^{B(\ep)} dx \,\, \glog(x)$. 
By construction the observed distribution functions are compatible with the  
maximum entropy principle with the usual constraints. 

By demonstrating that this entropy is compatible with the correct thermodynamics, 
we claim that it makes sense to talk about the thermodynamics of complex statistical systems.
This form of the entropy provides a potential tool that allows to carry out the usual thermodynamic 
operations in a fully self-consistent way. 

We have demonstrated that this entropy can be derived from the standard generalized 
Gibbs entropy ($\int B \glog B$) by adding a constant $\eta$ which is functionally 
depending on the measured distribution function \cite{lutsko}.
This $\eta$ is introduced to remove the problematic term $B \glog ' (B)$,  
arising from the variation of Gibbs entropy.
For the Boltzmann-Gibbs-Jaynes case, $\glog=\log$, and the problematic term is $B\glog'(B)=1$. 
This 1 can be absorbed into the constant $\gamma$, and the problematic term has vanished.  
Similarly, for Tsallis distributions we have $\glog=\ln_q$ and $B\glog'(B)=1+(1-q)\ln_q$. 
The 1 can get absorbed into $\gamma$, the $(1-q)\ln_q$  into the $q$-log term.  
In \cite{kaniadakislog} a scale-and-shift Ansatz was suggested to fix $B\glog'(B)=a\glog(B/b)+c$, 
where again terms can get absorbed. 
From this perspective our suggestion is different: We re-define the form of entropy to remove 
the term $B\glog'(B)$ altogether, not to absorb it.
As a consequence 
we end up with the most general entropy which is compatible with the maximum entropy 
principle under the usual constraints allowing arbitrary distribution functions. 
Further, the usual thermodynamic properties have been shown to hold \cite{abe07}.

$\eta$ allows for a physical interpretation of our philosophy: it somehow captures the numbers 
of states in phase space which depart from the classical Boltzmann case. This number 
may depend on long-range interactions or also parameters like temperature. 
The functional form of measured distributions, which is a kind of knowledge about the system, 
is thus naturally fed into the definition of the entropy of the (complex) statistical system. 
Effectively and formally, our result amounts to  replacing the $p\ln p$ term in the usual entropy by 
the integral, $\int \glog(p)$.
Obviously, classical Boltzmann-Gibbs entropy is obtained for specifying $\glog(x)=\ln(x)$; 
Tsallis entropy is a further natural special case.
Moreover, in this framework it can be easily shown that generalized distribution functions 
naturally imply the occurrence of dual logarithms. 

A further detail in our proposed entropy definition is that it does not contain 
any additional parameters, once the distribution is known. Once given the 
experimental distributions, there is no more freedom of choice of  generalized logarithms, 
nor of the functional form of the constant $\eta$. 

 S.T. would like to thank Andrea Rapisarda for the kind invitation to Catania.

%


\bibliographystyle{aipproc}   

\end{document}